\documentclass[twocolumn]{aastex61}

\begin{document}

\newcommand{\kms}{km~s$^{-1}$}
\newcommand{\msun}{$M_{\odot}$}
\newcommand{\rsun}{$R_{\odot}$}
\newcommand{\teff}{$T_{\rm eff}$}
\newcommand{\logg}{$\log{g}$}
\newcommand{\mas}{mas~yr$^{-1}$}

\accepted{August 16, 2017}

\title{Discovery of a Detached, Eclipsing 40 min Period Double White Dwarf 
Binary and a Friend:  Implications for He+CO White Dwarf Mergers\footnote{
	Based on observations obtained at the MMT Observatory, a joint facility of 
the Smithsonian Institution and the University of Arizona, and on observations 
obtained with the Apache Point Observatory 3.5-meter telescope, which is owned and 
operated by the Astrophysical Research Consortium.} }

\author{Warren R.\ Brown} \affiliation{Smithsonian Astrophysical Observatory, 60
Garden St, Cambridge, MA 02138 USA}

\author{Mukremin Kilic} \affiliation{Homer L. Dodge Department of Physics and
Astronomy, University of Oklahoma, 440 W. Brooks St., Norman, OK, 73019 USA}

\author{Alekzander Kosakowski} \affiliation{Homer L. Dodge Department of Physics and
Astronomy, University of Oklahoma, 440 W. Brooks St., Norman, OK, 73019 USA}

\author{A.\ Gianninas} \affiliation{Homer L. Dodge Department of Physics and
Astronomy, University of Oklahoma, 440 W. Brooks St., Norman, OK, 73019 USA}

\email{wbrown@cfa.harvard.edu, kilic@ou.edu, alexg@nhn.ou.edu}

\shorttitle{ Detached Eclipsing Double White Dwarf Binary and Friend }
\shortauthors{Brown et al.}

\begin{abstract}

	We report the discovery of two detached double white dwarf (WD) binaries, 
SDSS J082239.546+304857.19 and SDSS J104336.275+055149.90, with orbital periods of 
40 and 46 min, respectively.  The 40 min system is eclipsing; it is composed of a 
0.30 \msun\ and a 0.52 \msun\ WD.  The 46 min system is a likely {\it LISA} 
verification binary. The short $20\pm2$ Myr and $\sim$34 Myr gravitational wave 
merger times of the two binaries imply that many more such systems have formed and 
merged over the age of the Milky Way.  We update the estimated Milky Way He+CO WD 
binary merger rate and affirm our previously published result: He+CO WD binaries 
merge at a rate at least 40 times greater than the formation rate of stable 
mass-transfer AM~CVn binaries, and so the majority must have unstable mass-transfer.  
The implication is that spin-orbit coupling in He+CO WD mergers is weak, or perhaps 
nova-like outbursts drive He+CO WDs into merger as proposed by Shen.

\end{abstract}


\section{INTRODUCTION}

	There are about 100 double degenerate white dwarf (WDs) binaries known with 
orbital periods less than about 1 d \citep[e.g.][]{saffer98, bragaglia90, marsh95, 
moran97, maxted00, moralesrueda05, nelemans05b, vennes11, brown16a, rebassa17, 
breedt17, kilic17}.  Short-period WD binaries with periods less than 1~h have 
gravitational wave merger times less than about 100 Myr, and thus are interesting 
gravitational wave sources at mHz frequencies \citep{nelemans09, marsh11, 
nissanke12}.  The 765 sec orbital period binary J0651 \citep{brown11b}, for example, 
should be detected by the proposed {\it LISA} gravitational wave detector shortly 
after it is turned on \citep{korol17}.
	Short-period WD binaries are also interesting because they must either
evolve into stable AM~CVn systems, explode as supernovae, or merge into single
massive WDs, R~CrB stars, and related objects \citep[e.g.][]{webbink84, iben90}.  
None of these transformations have been observed directly, but we can compare
WD merger rates with different formation rates to constrain their outcome.

	Here, we report the discovery of two detached, double WD binaries, SDSS 
J082239.546+304857.19 and SDSS J104336.275+055149.90, with orbital periods of 40 and 
46 min, respectively.  We will henceforth refer to these objects as J0822 and J1043.  
J0822 and J1043 are the 5th- and 6th-shortest period WD binaries discovered by our 
Extremely Low Mass (ELM) Survey, a targeted spectroscopic survey for extremely low 
mass $\approx$0.2~\msun\ He-core WDs \citep[and references therein]{brown16a}.  
Thus we refer to degenerate $\approx$0.2~\msun\ objects as ELM WDs. Practically all 
ELM WDs are observed in compact binaries, with typical $M_2=0.76\pm0.25$ \msun\ WD 
companions and median $P=5.5$~h periods \citep{brown16a}.  J0822 and J1043 bring our 
ELM Survey sample to 82 binaries, more than half of which have merger times less 
than a Hubble time.

	Eclipses provide accurate constraints on the physical parameters of 
binaries.  J0822 is the seventh eclipsing double WD binary known after NLTT~11748 
\citep{steinfadt10}, CSS~41177 \citep{drake10, parsons11}, GALEX~J1717 
\citep{vennes11}, SDSS~J0651 \citep{brown11b}, SDSS~J0751 \citep{kilic14b}, and 
SDSS~J1152 \citep{hallakoun16}.  J0822 has a total mass of 0.82 \msun, a mass ratio 
of about 1:2, and will merge in 20 Myr.  

	The existence of double WD binaries with $\sim$10 Myr merger times implies 
that many more such systems have formed and evolved over the age of the Milky Way.  
Conversely, longer period systems remain binaries for the age of the Milky Way and 
must accumulate in observed samples.  An obvious question is what merging ELM WD 
binaries like J0822 and J1043 become.

	Previously, we used the magnitude-limited ELM Survey to estimate the local 
space density of ELM WD binaries and then calculate their merger rate by 1) 
inverting the distribution of merger times and 2) by forward-modeling different 
trial distributions to match the observations.  The major source of uncertainty 
comes from the small number statistics of rapidly merging binaries.  The 
gravitational wave merger timescale depends most strongly on orbital period 
\citep{kraft62}, so the shortest-period binaries dominate the merger rate estimate.  
Given that J0822 and J1043 increase the sample of $P<1$~h ELM WD binaries by 40\%, 
we revisit the merger rate estimate in light of the new discoveries.

	We begin by presenting our spectroscopic and photometric observations of 
J0822 and J1043.  We fit stellar atmosphere models to the spectra, orbital 
parameters to the radial velocities, and light curve parameters to the photometry.  
The Galactic kinematics of the binaries suggest that J0822 is a halo object while 
J1043 is a thin disk object.  We discuss the mass and mass ratio of J0822 in the 
context of other eclipsing double WD binaries, and close with an update on the 
merger rate of He+CO WD binaries in the Milky Way.

\section{DATA}

\subsection{Target Selection}

	The ELM Survey is a spectroscopic survey of low mass WD candidates selected 
on the basis of broadband color \citep{brown12a}.  We have also targeted some 
objects on the basis of stellar atmosphere fits to pre-existing Sloan Digital Sky 
(SDSS) spectra \citep{kilic10, kilic11a, kilic14}.  J0822 is an example of an object 
identified from its SDSS spectrum.  In 2016 February we obtained a pair of spectra 
to validate its nature and test for radial velocity variability.  We followed-up 
J0822 with time-series spectroscopy in 2016 October and 2017 March to determine its 
orbit.

	We targeted J1043 as part of the main ELM Survey \citep{brown12a}.  A single 
spectrum obtained in 2014 April identified J1043 as a likely low mass WD; a pair of 
spectra obtained in 2016 February revealed J1043 is velocity variable.  We 
re-observed J1043 with time-series spectroscopy in 2016 December and 2017 March.  
The year-long observing time baseline for both objects provide strong orbital 
period constraints using radial velocity alone.

\subsection{Spectroscopy}

	We obtain spectra in the same way as described in previous ELM Survey 
papers.  In brief, we use the 6.5m MMT telescope Blue Channel spectrograph with the 
832 l~mm$^{-1}$ grating in 2nd order, providing us with 1 \AA\ spectral resolution 
over $3600 < \lambda < 4500$ \AA.  We pair all spectra with comparison lamp 
exposures for accurate wavelength calibration.  We measure radial velocities with 
the cross-correlation package RVSAO \citep{kurtz98}.  We adjust exposure times to 
observing conditions, and obtain median 35 \kms\ errors for J0822 using exposure 
times between 12 and 22 min.  J1043 is a brighter target, and we obtain median 20
\kms\ radial velocity errors using exposures times between 6.5 and 9 min.  We became 
concerned about the length of our exposure times after we discovered the velocities 
were phasing at $\approx$45 min periods.

	To properly sample the phase curves and validate the short orbital periods, 
we obtained 1.5~h worth of back-to-back spectra for both objects at 2 \AA\ 
resolution (with the MMT Blue Channel spectrograph 800 l~mm$^{-1}$ grating in 1st 
order) using half the normal exposure times.  The radial velocity uncertainties are 
worse with this set-up, however the time series confirm the $\approx$45 min periods.

	In total, we obtained 20 spectra of J0822 and 31 spectra of J1043.  We 
provide the radial velocities in a Data behind the Figure table accessible in the 
on-line journal.  Both objects are observed to be single-lined spectroscopic 
binaries, as seen in Figure \ref{fig:spec}.

\subsection{High-Speed Photometry}

	We obtained time-series photometry for J0822 and J1043 using the Apache 
Point Observatory 3.5m telescope with the Agile frame-transfer camera 
\citep{mukadam11} and the BG40 filter on the night of UT 2 March 2017.  We use 
exposure times of 30~s for J0822 and 15~s for J1043, and obtain total integrations 
of 68 min and 60 min, respectively. There is a 5.6 min gap in the J0822 data due to 
an instrument problem. 

	We use standard IRAF\footnote{IRAF is distributed by the National Optical 
Astronomy Observatories, which are operated by the Association of Universities for 
Research in Astronomy, Inc., under cooperative agreement with the National Science 
Foundation.} routines to perform aperture photometry.  We correct for transparency 
variations using two relatively bright comparison stars in the field of view of each 
WD.  After detecting eclipses in J0822, we attempted additional follow-up on three 
different nights.  All follow-up attempts were unfortunately lost to weather.

\section{ANALYSIS AND RESULTS}

\subsection{Stellar Atmosphere Fits}

	We perform stellar atmosphere fits in the same way as described in previous 
ELM Survey papers.  In brief, we fit the summed, rest-frame spectra to a grid of 
pure hydrogen atmosphere models that span 4,$000~{\rm K} < T_{\rm eff} < 35$,000~K 
and $4.5 < \log{g} < 9.5$ \citep{gianninas11, gianninas14b, gianninas15} and that 
include the Stark broadening profiles from \citet{tremblay09}.  J1043 has a 
temperature and gravity that require three-dimensional stellar atmosphere model 
corrections, which reduce the 1D parameters by 230 K and 0.23 dex 
\citep{tremblay15}.  We present the corrected parameters in Table \ref{tab:param}.

\begin{figure} 
 \plotone{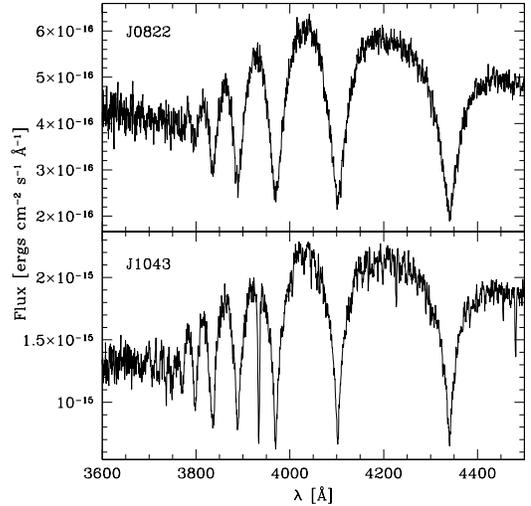}
 \caption{ \label{fig:spec}
	Summed rest-frame spectra of J0822 (top panel) and J1043 (bottom panel).
} \end{figure}

	J1043 is also a DAZ WD that exhibits strong Ca~{\sc ii} $\lambda$3933 and
Mg~{\sc ii} $\lambda$4481 absorption lines (Figure \ref{fig:spec}).  The Ca~{\sc ii}
and Mg~{\sc ii} lines show the same radial velocity variability as measured from the
Balmer lines, and so must come from the WD.  We and others regularly see Ca~{\sc ii}
in other $\log{g}\sim6$ ELM WD spectra \citep{brown13a, kaplan13}.  We mask the
region around Ca {\sc ii} when doing the Balmer line fits.

\begin{deluxetable}{lll}	
\tablecolumns{3}
\tablewidth{0pt}
\tablecaption{System Parameters\label{tab:param}}
\tablehead{ \colhead{Parameter} & \colhead{J0822} & \colhead{J1043} }
        \startdata
RA (J2000)	& 8:22:39.546		& 10:43:36.275		\\
Dec (J2000)	& +30:48:57.19		& +5:51:49.90		\\
\teff\ (K)	& $13920 \pm 255$	& $9260 \pm 140$	\\
\logg\  	& $7.14 \pm 0.05$	& $6.60 \pm 0.06$	\\
$M_1$ (\msun)	& $0.304 \pm 0.014$	& $0.183 \pm 0.010$	\\
$M_g$ (mag)	& $+9.96 \pm 0.09$	& $+10.23 \pm 0.11$	\\
$g_0$ (mag)	& $20.198 \pm 0.023$	& $19.054 \pm 0.017$	\\
d$_{\rm helio}$ (kpc) &	$1.11 \pm 0.08$	& $0.58 \pm 0.05$	\\
$P$ (d) 	& $0.02797 \pm 0.00016$	& $0.03170 \pm 0.00092$	\\
$k$ (\kms)	& $415.7 \pm 22.7$	& $115.2 \pm 6.8$	\\
$\gamma$ (\kms)	& $10.2 \pm 23.0$	& $31.7 \pm 4.6$	\\
$M_2$ (\msun)	& $0.524 \pm 0.050$	& $>0.07$		\\
		&			& (if $0.76 \pm 0.25$)	\\
$i$ (deg)	& $88.1^{+1.4}_{-2.3}$	& $<85.7$ 		\\
		&			& (then $12.6^{+2.9}_{-1.8}$)\\
$a$ (\rsun)	& $0.364 \pm 0.008$	& $>0.27$		\\
		&			& (and $0.41 \pm 0.04$)	\\
$\tau_{\rm merge}$ (Myr) & $20 \pm 2$ 	& $<240$		\\
		&			& (and $34^{+12}_{-7}$)	\\
$\log{h}$	& $-22.36 \pm 0.05$	& $>-22.9$		\\
		&			& (and $-21.7 \pm 0.1$)	\\
$R_1$ (\rsun)	& $0.022^{+0.015}_{-0.014}$	& \nodata	\\
$R_2$ (\rsun)	& $0.0041^{+0.0082}_{-0.0037}$	& \nodata	\\
$L_2/L_1$	& $0.00^{+0.04}_{-0}$		& \nodata	\\
	\enddata
\end{deluxetable}

\begin{figure*}		
 \plottwo{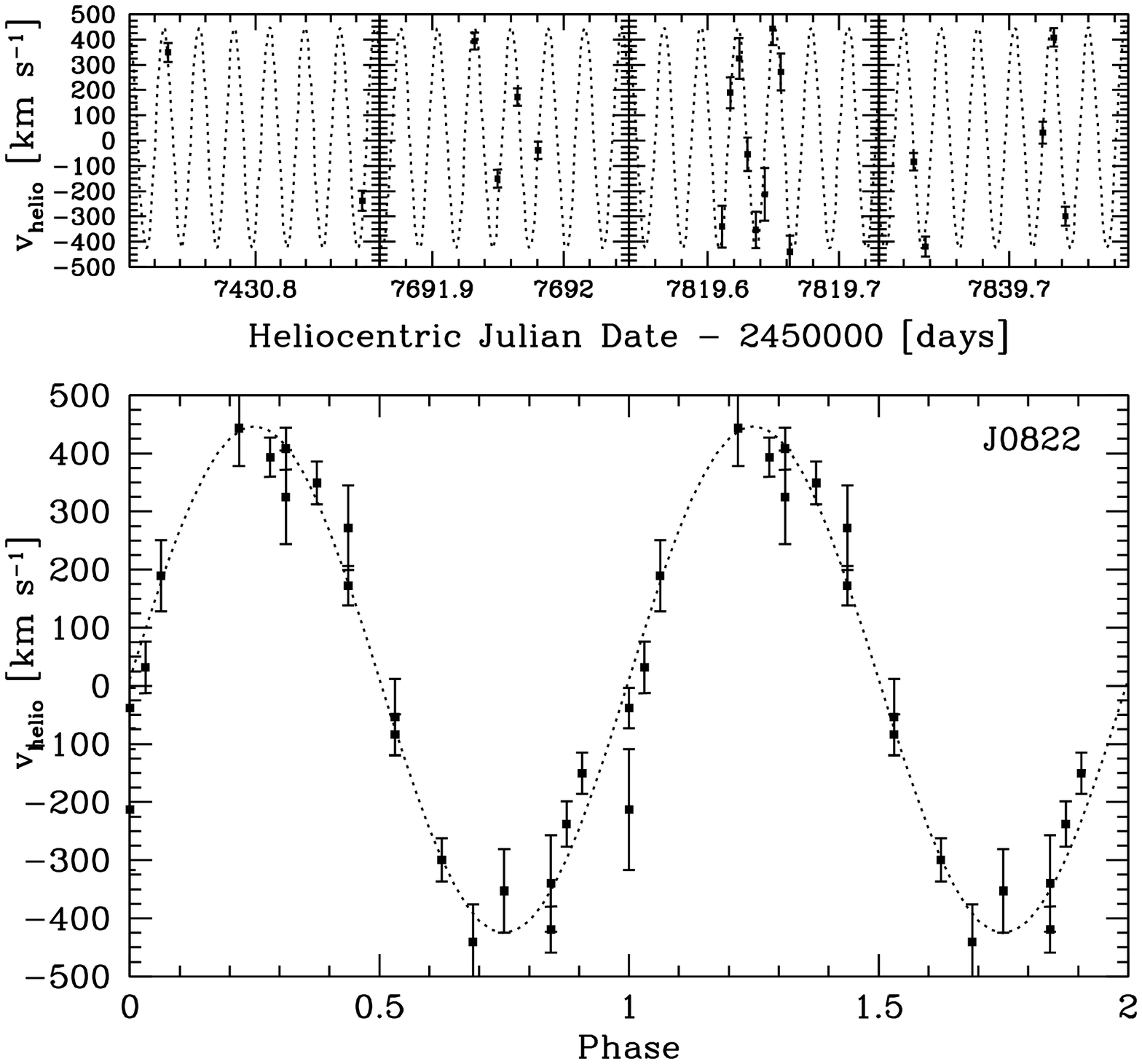}{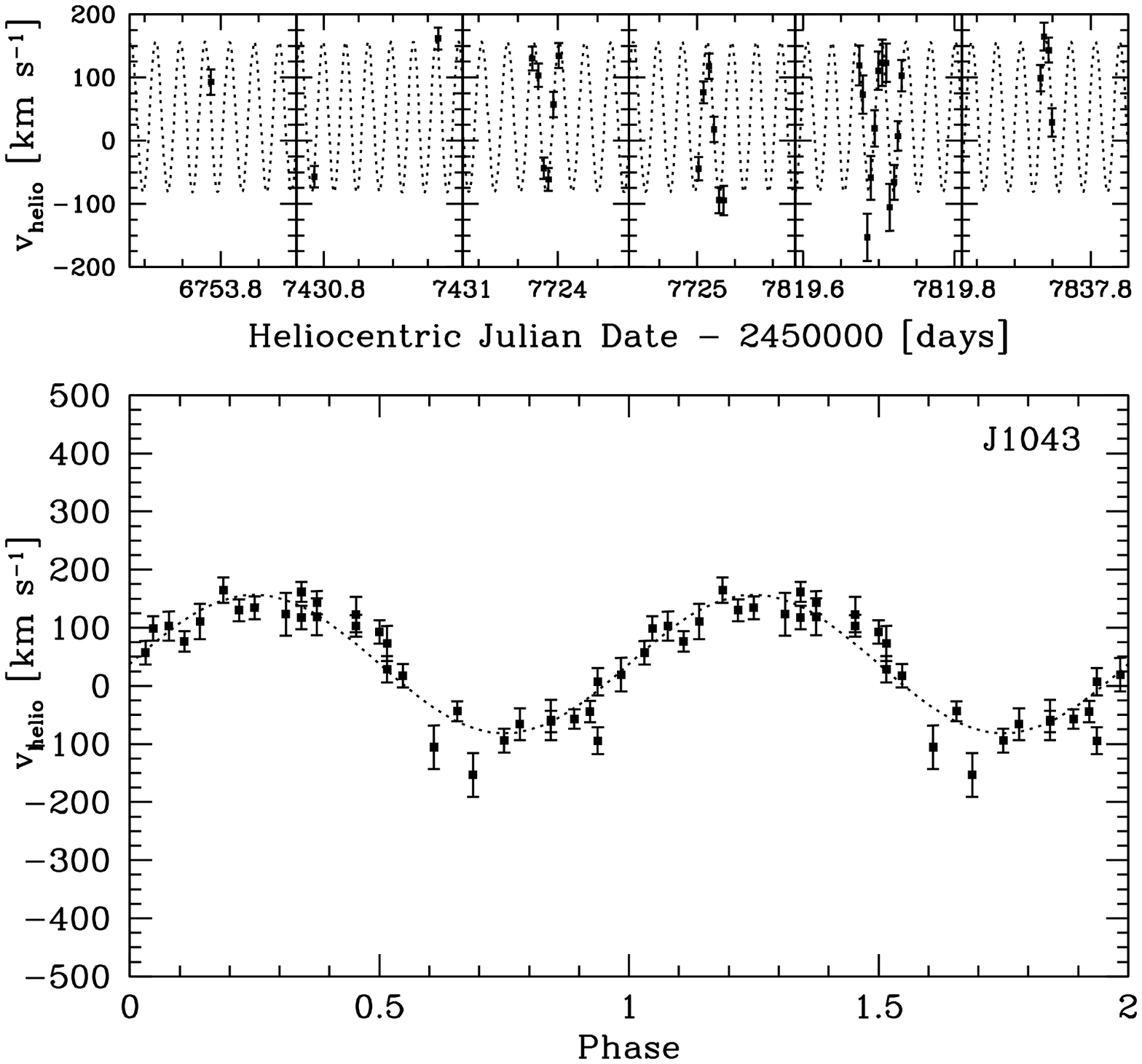}
 \caption{ \label{fig:rv}
	Radial velocity observations of J0822 and J1043.  Nightly measurements are 
plotted in the small panels; measurements phased to the $\chi^2$ minima at 
$P=0.028126$ d and $P=0.032618$ d, respectively, are plotted in the large panels. The radial velocities are available as the Data behind the Figure.
} \end{figure*}

\subsection{White Dwarf Parameters}

	We estimate WD mass and luminosity by matching the measured \teff\ and 
\logg\ to the ELM WD evolutionary tracks of \citet{istrate16}.  The tracks account 
for the effects of element diffusion and rotation mixing, beginning at the moment 
the progenitor detaches from the mass-transfer phase.  Progenitor metallicity can 
have a significant impact on the hydrogen envelope mass, the number of thermonuclear 
shell flashes, and the resulting cooling time of the tracks \citep{althaus15}.  
Thus we apply $Z=0.02$ tracks to disk objects, and $Z=0.001$ tracks to halo objects.  
As a cross-check, we compare with \citet{althaus13} solar metallicity tracks.

	We interpolate the evolutionary tracks in the same way as described in 
previous ELM Survey papers.  Our approach is to identify the two nearest tracks to 
an observed \teff\ and \logg\ value, and interpolate between the tracks on the basis 
of \logg .  Loops due to shell flashes complicate the picture, and so we re-sample 
\teff\ and \logg\ with their errors to estimate the dispersion of the mass and 
luminosity estimates.  The results are presented in Table \ref{tab:param}.

	J0822 has mass $M_1 = 0.304 \pm 0.014$ \msun\ and absolute $g$-band 
magnitude $M_g = +9.96 \pm 0.09$ based on \citet{istrate16} $Z=0.001$ tracks.  With 
a de-reddened apparent magnitude $g_0 = 20.198 \pm 0.023$, J0822 is 1.11 kpc distant 
and likely a halo object (see below).

	J1043, on the other hand, has $M_1 = 0.183 \pm 0.010$ \msun\ and $M_g = 
+10.23 \pm 0.11$ mag based on \citet{istrate16} $Z=0.02$ tracks.  With a de-reddened 
apparent magnitude $g_0 = 19.054 \pm 0.017$, J1043 is 0.58 kpc distant and likely a 
disk object (see below).
	Mass and luminosity estimates from \citet{althaus13} tracks agree to within 
1-$\sigma$ for both objects, thus the mass and luminosity estimates appear robust to 
the choice of evolutionary tracks.

	Interestingly, the evolutionary tracks predict that J1043 has undergone 
multiple thermonuclear shell flashes, while J0822 has not.  \citet{istrate16} argue 
that rotational mixing can keep metals visible at the surface of a WD for 
longer-than-expected periods of time after shell flashes.  This prediction is 
consistent with the strong metal lines present in J1043 and absent in J0822.

\subsection{Disk / Halo Kinematics}

	We use kinematics to determine whether the objects belong to the disk or 
halo.  The gravitational redshift corrections for J0822 and J1043 are $7.9\pm0.3$ 
and $3.3\pm0.2$ \kms, respectively.  The objects have comparable systemic radial 
velocities, $\gamma_{\rm J0822}=10.2 \pm 23.0$ \kms\ and $\gamma_{\rm J1043}=31.7 
\pm 4.6$ \kms, but different proper motions.

	We obtain proper motions from the HSOY catalog \citep{altmann17}, a new 
proper motion catalog that uses {\it Gaia} Data Release 1 positions for its final 
epoch.  While J0822 is not detected in all epochs, its proper motion $\mu_{\rm 
J0822}=29.7 \pm 7.9$ \mas\ is formally significant.  J1043 is detected in all epochs 
and has a smaller proper motion, $\mu_{\rm J1043} = 8.5 \pm 3.6$ \mas.

	We calculate velocities in the Galactic rest frame assuming a circular 
velocity of 235 \kms\ and the Local Standard of Rest motion of \citet{schonrich10}.  
We use \citet{chiba00} velocity ellipsoid values to put the motions in context.  
J0822's space motion, $(U,V,W) = (8\pm25, -141\pm31, -29\pm28)$ \kms, falls outside 
the 2-$\sigma$ velocity dispersion threshold of the thick disk, but lies well within 
the 1-$\sigma$ velocity dispersion threshold of the halo.  On this basis, we 
identify J0822 as a likely halo object.  J1043's space motion, $(U,V,W) = (-20\pm7, 
-6\pm7, 24\pm6$) \kms, is consistent with the disk, and so we consider J1043 a 
likely disk object.

\subsection{Binary Orbital Elements}

	We calculate orbital elements in a similar way as described in previous ELM 
Survey papers.  In brief, we minimize $\chi^2$ for a circular orbit following the 
code of \citet{kenyon86}.  To compare the model to the observations, we average the 
model over each exposure time as we search through period and phase.  The 
timebaseline and phase coverage of the observations (Figure \ref{fig:rv}) yield a 
well-defined $\chi^2$ minimum in both objects.  We determine the best-fit parameters 
from the envelope of the $\chi^2$ minima, which are symmetric but have sub-structure 
due to our sampling.  The orbital fits to J0822 and J1043 have reduced $\chi^2$ of 
1.06 and 1.17, respectively.

	We estimate errors by re-sampling the velocities with their errors and 
re-fitting the orbital parameters 10,000 times.  This Monte Carlo approach samples 
the $\chi^2$ space in a self-consistent way.  We report the median orbital 
parameters along with the averaged 15.9\% and 84.1\% percentiles of the 
distributions in Table \ref{tab:param}.  The systemic velocities are corrected for 
gravitational redshift.

	Kepler's 3rd law, written as the binary mass function, relates orbital 
period $P$, semi-amplitude $k$, ELM WD mass $M_1$, companion mass $M_2$, and 
orbital inclination $i$ as follows:
	\begin{equation} \frac{P k^3}{2 \pi G} = \frac{(M_2 
\sin{i})^3}{(M_1+M_2)^2}.\end{equation}
	We directly measure $P$ and $k$.  We derive $M_1$ from the observed \teff\
and \logg.  Given a constraint on $i$ (i.e.\ from eclipses or another observational
constraint), we can derive $M_2$.

	J0822 has $P=40.28 \pm 0.23$ min and semi-amplitude $k=415.7 \pm 22.7$ \kms.  
Assuming inclination $i=88.1^{+1.4}_{-2.3}$ deg (see below), Kepler's 3rd Law 
tells us that J0822's unseen companion has mass $M_2 = 0.524 \pm 0.05$ \msun\ and 
orbital separation $a = 0.364 \pm 0.008 $ \rsun.  J0822 is a double WD binary.  The 
gravitational wave merger time of the binary is remarkably short, $\tau = 20 \pm 2$ 
Myr, however the gravitational wave strain is only $\log{h} = -22.36 \pm 0.05$ given 
the distance and masses involved.  The eclipse light curve allows us to further 
characterize J0822 below.

	J1043 has $P=45.65 \pm 1.32$ min and $k=115.2 \pm 6.8$ \kms.  The larger 
period uncertainty reflects the broader envelope of its $\chi^2$ minimum.  In 
the absence of a constraint on inclination, Kepler's 3rd Law tells us that J1043's 
unseen companion must have mass $M_2 > 0.07$ \msun\ and orbital separation $a > 
0.27$ \rsun.  However, we can rule out a low-mass companion on observational 
and physical grounds.

	The radial velocities, taken alone, allow for a low mass M dwarf or perhaps 
a brown dwarf companion, like the recently discovered eclipsing system J1205$-$0242 
\citep{parsons17, rappaport17}.  However, an M dwarf should fill the Roche lobe at 
this orbital separation.  We see no evidence of mass transfer.  An M dwarf should 
also out-shine the WD at infrared wavelengths.  We compare publically available {\it 
GALEX} ultraviolet, SDSS optical, and UKIDSS infrared photometry to the synthetic WD 
spectral energy distribution and find good agreement with the WD model, and no 
evidence for infrared excess.  The close orbital separation and lack of evidence for 
an M dwarf companion suggest that J1043 is a double degenerate binary.

\subsubsection{Gravitational Wave Detection}

	Double degenerate binaries like J1043 are mHz sources of gravitational 
waves.  The strongest sources of gravitational waves in the mHz frequency range will 
be directly detected by {\it LISA}.  We use the detection calculations of 
\citet{korol17} to estimate the signal-to-noise ratio (SNR) at which J1043 might be 
detected by {\it LISA}.  The signal amplitude is proportional to the gravitational 
wave strain of a binary times the {\it LISA} detector pattern function.  
J1630+4233, a previously discovered ELM WD binary that has an orbital frequency 
nearly identical to J1043 \citep{kilic11c}, provides an appropriate comparison.  
\citet{korol17} predict that {\it LISA} will detect J1630+4233 at SNR=5 in 5 years 
of operation.

	J1043's gravitational wave strain is identical to J1630+4233 if J1043 has 
$M_2=0.38$ \msun.  Thus {\it LISA} will detect J1043 at SNR=5 in 5 years of 
operation if its mass ratio is $M_1$:$M_2$=1:2.  Re-calculating gravitational wave 
strain for different mass ratios, {\it LISA} will detect J1043 at SNR=3 if its mass 
ratio is 1:1, and at SNR=9 if its mass ratio is 1:5.  J1043 is a detectable 
gravitational wave source for even the most pessimistic 1:1 mass ratio.

	Although we do not measure the mass ratio, we expect that a degenerate 
companion should be more massive than the observed low mass WD.  ELM WDs are 
understood to be the result of double common-envelope evolution, in which the ELM WD 
evolves last \citep{webbink84, iben90, marsh95}.  The Universe is not old enough to 
evolve a single star into an ELM WD.  Indeed, eclipsing binaries in the ELM Survey 
with well-determined parameters have measured mass ratios between $M_1$:$M_2$=1:2 
and 1:5 \citep{brown11b, kilic14b}.

	We conclude that J1043 is a likely {\it LISA} verification binary.  To 
simplify the remaining discussion, we will assume that J1043's companion has the 
average mass found in the rest of the ELM Survey, $M_2 = 0.76 \pm 0.25$ \msun\ 
\citep{andrews14, boffin15, brown16a}, and thus a mass ratio of $M_1$:$M_2$=1:4.  
This choice of $M_2$ corresponds to an orbital inclination of $i = 
12.6_{-1.8}^{+2.9}$ deg, a gravitational merger time of $\tau = 34_{-7}^{+12}$ Myr, 
and a gravitational wave strain of $\log{h} =-21.7 \pm 0.1$.  Future gravitational 
wave measurements will tell us the exact answer.

\begin{figure}          
 \includegraphics[angle=270, width=3.25in]{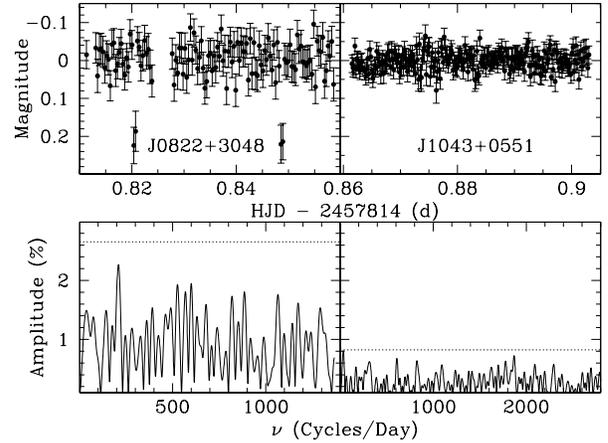}
 \caption{ \label{fig:lc}
	Light curves of J0822 and J1043 (upper panels) and their Fourier transforms 
(lower panels).  Dotted lines are the 3$<$A$>$ significance limits as defined in the 
text.  The relative photometry is available as the Data behind the Figure.  J0822 
shows 0.2 mag deep eclipses with a period identical to the radial velocity data.} 
\end{figure}

\subsection{Eclipse Light Curve}

	Figure \ref{fig:lc} presents the light curves for J0822 and J1043. J0822 
shows 0.2~mag deep and $\approx$60~s long eclipses every 40.5 min, statistically 
identical to the orbital period derived from the radial velocity data.
	We do not detect the secondary eclipse.  J0822 is faint, with apparent 
$g=20.34$ mag, and our photometry is relatively noisy, with $\pm$0.04 mag errors.  
Because we have only 4 data points during the primary eclipse, the Fourier transform 
of J0822 (bottom panel) does not show significant variability above the 3$<$A$>$ 
limit, where $<$A$>$ is the average amplitude up to the Nyquist frequency.  The 
relativistic beaming effect should produce 0.35\% variations at the orbital period 
\citep{shporer10}, but this is also lost in the noise in our data.  Tidal 
distortions are not significant in J0822.  The oblateness of the low mass WD is 
predicted to be 0.3\% and the predicted amplitude of the ellipsoidal variations is 
$\sim10^{-4}$, which we are unlikely to detect with ground-based observations.

	The light curve and the Fourier transform for J1043 do not show any 
significant variability down to the 3$<$A$>$ limit of 0.8\%.  The absence of 
eclipses provides us with an upper limit on the inclination of the system, 
$i<85.7^{\circ}$. Given the unknown inclination, the expected amplitudes of the 
relativistic beaming effect and ellipsoidal variations are $\leq0.3$\% and 
$\leq0.15$\%, respectively. If present, these photometric signals are also lost in 
the noise in our observations.

	We model the light curve of J0822 using JKTEBOP \citep{southworth04}. We use 
the wavelength response of the BG40 filter from \citet{hallakoun16} and the linear 
limb-darkening coefficients from \citet{gianninas13} to calculate the limb-darkening 
coefficients. We adopt gravity-darkening coefficients of 0.36 for both the primary 
and secondary star. We expect convection to be present in both stars, so adopting 
$\beta = 0.36$ is reasonable. Given how sparsely our data sample the primary 
eclipse, we avoid using more complicated limb-darkening laws. The orbital period and 
the mass ratio of the system are well-constrained by the radial velocity data.  
We thus fix the orbital period, mass ratio, limb-, and gravity-darkening 
coefficients when fitting for the inclination and component radii.

\begin{figure}          
 \includegraphics[angle=270, width=3.25in]{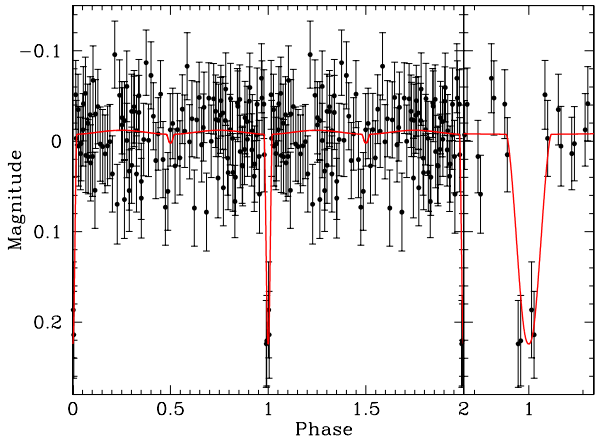}
 \caption{ \label{fig:eclipse}
	Light curve model of J0822 (red line) compared to the data (black dots).  
Orbital period and mass ratio are better constrained by the radial velocity data, so 
we hold those parameters fixed; the derived WD radii are consistent with model 
predictions for 0.3 and 0.52 \msun\ WDs.
} \end{figure}

	Figure \ref{fig:eclipse} shows the best-fit model, which has an inclination
of $86.8^{\circ}$ and reduced $\chi^2=0.96$.  We perform 10,000 Monte Carlo
simulations \citep{southworth05} to estimate the errors in each parameter. Because
we only have 4 data points during the primary eclipse, about half of the Monte Carlo
simulations find a best-fit solution with $i<84.5^{\circ}$.  This is ruled out by
the detected eclipses.  Treating the radii as free parameters, a 0.22 mag
eclipse depth requires $i\ge84.5^{\circ}$ for J0822.  Below $i<84.5^{\circ}$, the
model eclipse depths become too small for the observed light curve.  Thus we limit
the Monte Carlo simulations to the solutions with $i\geq84.5^{\circ}$ when
estimating errors.  The median parameters with their errors are:  
$i=88.1_{-2.3}^{+1.4}$ deg, $R_1 = 0.022_{-0.014}^{+0.015} R_{\odot}$, $R_2 =
0.0041_{-0.0037}^{+0.0082} R_{\odot}$, and $L_2 / L_1 = 0.000_{-0}^{+0.038}$.  
Using the period determined from the radial velocity observations, the best-fit
ephemeris is $T_0 = 2457814.821 \pm 0.007$ BJD$_{\rm TDB}$. Note that gravitational
lensing minimally affects the depth of the primary and secondary eclipses, by
$<0.6$\% \citep{marsh01}, and has not been included in our modeling.

	A better light curve is, of course, desirable, but the derived parameters 
remain consistent with model predictions for 0.30 and 0.52 \msun\ WDs.  The 
eclipsing WD binary CSS~41177 contains a $0.316\pm0.011$ \msun\ WD with $R=0.02066 
\pm 0.00042$ \rsun \citep{bours14}, very similar to J0822.  Since the secondary 
eclipse in J0822 is not detected in our observations, the luminosity of the more 
massive companion star is unconstrained, but it is likely a few per cent of the 
luminosity of the visible WD.  Multi-passband and higher signal-to-noise ratio 
follow-up photometry of J0822 will be useful for obtaining more precise constraints 
on the radii, luminosities, and temperatures of the two WDs.

	Eclipsing short period systems provide excellent clocks to detect orbital 
decay due to gravitational waves. \citet{hermes12c} measured \.{P}$ =-9.8 \pm 2.8 
\times 10^{-12}$ s~s$^{-1}$ in the 765~s eclipsing double WD binary J0651; the 
precision in this measurement has recently improved to better than $0.1 \times 
10^{-12}$ s s$^{-1}$ after 5 years of observations (J.J. Hermes, private 
communication).  The expected \.{P} for J0822 is about $-1.4 \times 10^{-12}$ 
s~s$^{-1}$ \citep{piro11}.  That means J0822's time-of-eclipse should occur 7~s 
earlier than expected in 5 years, and occur 30~s earlier than expected in 10 years.  
A reliable measurement of \.{P} for J0822 will thus likely require a 5 to 10 year 
time baseline.

\section{DISCUSSION}

\subsection{Eclipsing Double White Dwarf Binaries}

	There are now 7 detached, eclipsing double WD binaries known whose WD 
masses are directly measured.  An obvious question is what happens when these double 
WD binaries merge due to gravitational wave radiation and begin mass transfer.  
Total mass is important for whether the binaries are type Ia supernova progenitors; 
mass ratio is important for whether the binaries will undergo stable or unstable 
mass transfer \citep[e.g.][]{iben90, han98, nelemans01b, marsh04, kaplan12}.

\begin{figure}          
 \includegraphics[width=3.5in]{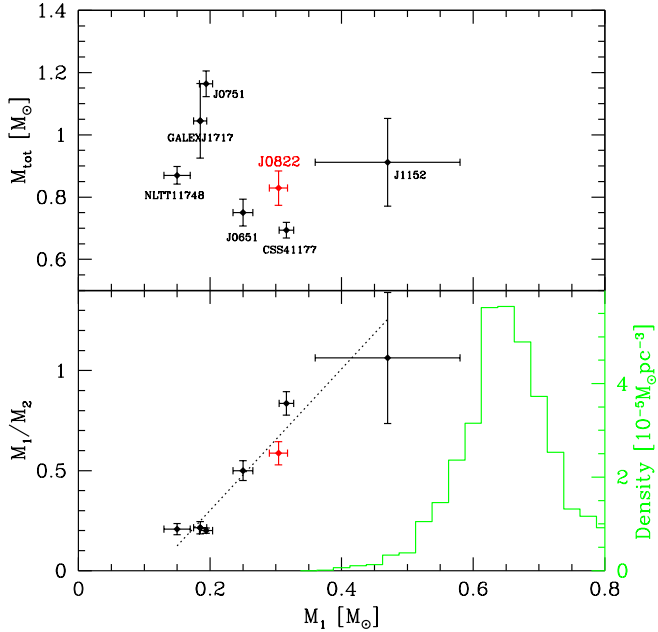}
 \caption{ \label{fig:mtotq}
	Total mass (upper panel) and mass ratio (lower panel) of known eclipsing 
double WD binaries, plotted versus mass of the most luminous WD.  Green histogram is 
the mass distribution of WDs observed in SDSS \citep{kepler16}.  None of the 
eclipsing WD binaries contain a 0.6 - 0.7 \msun\ WD of ``normal'' mass; a marginally 
significant correlation appears in mass ratio (dotted line). } \end{figure}

	Figure~\ref{fig:mtotq} plots the distribution of total mass and mass ratio 
of the 7 known detached, eclipsing WD binaries.  $M_1$ is the WD that dominates the 
light of the binary, which is typically the lowest mass WD in the binary.  Low mass 
WDs have larger radii and typically have higher temperatures (because they evolved 
more recently) than their higher mass WD companions.  The only double-lined 
spectroscopic binary among the eclipsing systems is the equal-mass binary CSS~41177.  
We obtain binary parameters from the following sources:  NLTT~11748 
\citep{kaplan14}, CSS~41177 \citep{bours14, bours15}, GALEX~J1717 \citep{vennes11, 
hermes14b}, SDSS~J0651 \citep{brown11b, hermes12c}, SDSS~J0751 \citep{kilic14b}, and 
SDSS~J1152 \citep{hallakoun16}.  We also draw (in green) the mass distribution of DA 
WDs measured from the SDSS DR12 WD catalog \citep{kepler16} for comparison.

	Curiously, none of the eclipsing double WD binaries contain a WD that falls 
in the peak of the WD mass distribution observed by SDSS (green histogram, Figure 
\ref{fig:mtotq}) \citep{kepler16}.  Rather, the eclipsing binaries all contain low 
mass $<0.5$ \msun\ WDs with companions that are either other low mass WDs, or else 
high mass $0.7 < M_2 < 1.0$ \msun\ WDs for the 1:5 mass ratio systems.  The total 
masses of the eclipsing double WD binaries range from $0.69\pm0.03$ \msun\ to 
$1.16\pm0.04$ \msun.

	The ubiquity of low mass WDs is not due to how the eclipsing double WD 
binaries were targeted.  NLTT~11748 was targeted for its large proper motion 
\citep{kawka09}.  CSS~41177 was identified in a search of Catalina Sky Survey light 
curves of $\sim$12,000 SDSS WDs \citep{drake10}.  GALEX~J1717 was found by its 
ultraviolet excess and reduced proper motion \citep{vennes11}.  SDSS~J1152 was a 
color-selected WD candidate included in the 2-Wheeled {\it Kepler} continuation 
mission \citep{howell14}.  Only the ELM Survey binaries were targeted for their low 
mass WD.  Yet because low mass WDs form in compact binaries and are typically more 
luminous than higher-mass WDs, binary population synthesis models predict that low 
mass WDs should dominate magnitude-limited samples of WD binaries 
\citep[e.g.][]{iben90, han98, nelemans01a}.

	However, the mass ratios of eclipsing double WD binaries are unexpected in 
view of binary population synthesis models.  Observed mass ratios range from 1:1 to 
1:5.  There also appears to be a marginally significant correlation of mass ratio 
(and thus $M_2$) with $M_1$, such that lower mass WDs have higher mass WD companions 
(as indicated by the dotted line in Figure \ref{fig:mtotq}).
	Yet models like \citet{han98} and \citet{nelemans01a} predict predominantly 
1:1.5 mass ratios for He+He WD binaries, and 1:2 to 1:3 mass ratios for He+CO WD 
binaries, in synthetic magnitude-limited samples of double WDs.  Binaries with 
$M_1\approx0.2$ \msun\ and extreme 1:5 mass ratios are missing from the models.  
Short-period WD binaries are sensitive to model assumptions about the stability of 
mass transfer and double common-envelope evolution \citep{toonen14}.  This has 
implications for the likely outcomes of the mergers.

	Binaries composed of He+CO WDs with mass ratios $M_1/M_2<0.2$ should 
experience stable mass transfer \citep{marsh04} and evolve into AM~CVn systems, a 
class of ultracompact binaries that consist of a WD accretor and a helium donor star 
\citep{nelemans05, solheim10}.  The three eclipsing binaries with $M_1 < 0.2$ \msun\ 
will likely evolve in this route.

	He+CO WD binaries with mass ratios greater than about $M_1/M_2>0.5$ should 
experience unstable mass transfer \citep{marsh04} and merge into single objects like 
R~CrB stars \citep{webbink84, iben90}.  The four eclipsing binaries with $M_1>0.25$ 
\msun\ will likely evolve into single $\sim$0.8 \msun\ objects.

\begin{figure}          
 \includegraphics[width=3.25in]{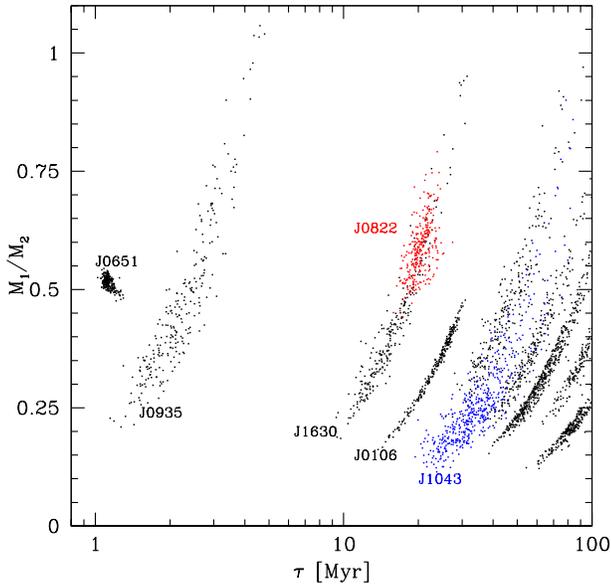}
 \caption{ \label{fig:tauq}
	Gravitational wave merger time versus mass ratio for double WD binaries in 
the ELM Survey with $\tau<100$ Myr.  Each dot represents a Monte Carlo calculation 
accounting for observational uncertainties plus constraints on inclination (i.e. 
eclipses).  J0822 is drawn in red, and J1043 is drawn in blue. } \end{figure}

\subsection{Merger Rate of Double WD Binaries}

	As we state above, finding double WD binaries that merge on $\sim$10 Myr 
timescales implies many more such binaries must have formed and merged over the age 
of the Milky Way.
	In \citet{brown16b}, we show that the distribution of ELM WD binaries in our 
magnitude-limited sample implies an ELM WD merger rate of $3\times10^{-3}$ yr$^{-1}$ 
in the Milky Way disk.  The major source of uncertainty comes from the small 
number statistics of rapidly merging binaries like J0822 and J1043.
	While J0822 formally falls outside the ELM Survey color and magnitude 
limits, J1043 is very much a part of the sample.  Assuming the prior on its 
secondary, J1043 ranks as the 5th shortest merger-time system in the ELM Survey 
after J0651, J0935, J1630, and J0106 (Figure \ref{fig:tauq}).

	We revisit the merger rate calculation and find that the addition of J1043 
increases the estimated merger rate of ELM WD binaries in the Galactic disk by only 
1\%.  The gain from its merger time is offset by the completeness correction (the 
sample is now more complete).  We conclude that our sample of ELM WDs is large 
enough to be robust to the addition of new $\tau\sim10$ Myr binaries; only the 
addition of $\tau\sim1$ Myr binaries like J0651 and J0935 can alter the rate at the 
10\% level.

	We also re-visit the merger rate of ELM WD binaries with extreme mass 
ratios, those systems that should evolve into stable mass-transfer AM~CVn.  J1043 is 
possibly one such a system.  Figure \ref{fig:tauq} illustrates the mass ratio 
constraints on ELM Survey binaries with merger times less than 100 Myr.  Every 
binary without an inclination measurement has a likelihood of having a mass ratio 
that falls in the stable mass-transfer regime.  We follow \citet{kilic16} and 
estimate the merger rate from the part of the distribution that meets the 
\citet{marsh04} criteria for stable mass transfer.  The uncertainties of this 
approach are large, but we obtain the same answer as previously published.  The 
estimated rate is 20 to 40 times lower than the full ELM WD merger rate, depending 
on how the estimate is made, consistent with the AM~CVn formation rate 
\citep{roelofs07b, carter13}.

	The ELM Survey samples only a subset of He+CO WD binaries, and so places
only a lower limit on the Milky Way's full He+CO WD merger rate.  The fact that an 
order-of-magnitude more short-period ELM WD binaries are merging than stable AM~CVn 
binaries are forming means that most ELM WD (and thus He+CO WD) binaries experience 
unstable mass transfer and merge into single objects.  Indeed, the ELM WD merger 
rate is statistically identical to the formation rate of R~CrB stars in the Milky 
Way \citep{zhang14, karakas15}.  

	From a theoretical viewpoint, the conclusion that most ELM WD binaries
experience unstable mass transfer is surprising.  Unstable mass transfer is only
expected for near-unity mass ratios.  However \citet{marsh04} show that there
remains a large region of parameter space in which the stability of mass transfer is
ambiguous, in which stability depends primarily on the strength of spin-orbit
coupling.  When the donor ELM WD fills its Roche lobe, the accretor spins-up due to
the incoming matter stream.  Weak spin-orbit coupling means the accretor is unable
to transfer angular momentum back to the orbit of the donor on a fast enough
timescale.  The result is that the binary orbit shrinks, mass transfer rate grows,
and mass transfer becomes unstable.  Alternatively, the initial phase of hydrogen
mass-transfer may generate nova-like outbursts that drive He+CO WD systems into
merger \citep{shen15}.  Either way, the merger rate of observed short-period ELM WD
binaries like J1043 demands that mass transfer in most He+CO WD binaries is
unstable.

\section{CONCLUSIONS}

	We present the discovery of a detached eclipsing 40 min orbital period
double WD binary, and a detached 46 min orbital period double WD binary.  These two
systems bring our targeted ELM Survey sample to 82 WD binaries.  The two new
systems, SDSS J082239.546+304857.19 and SDSS J104336.275+055149.90, have
gravitational wave merger times of 20 Myr and $\sim$34 Myr, respectively.  J0822 is
a detached, eclipsing double WD binary.  J1043 is a likely gravitational wave
verification binary.

	We revisit the ELM WD binary merger rate, and find that the new discoveries 
affirm our previous result: observed He+CO WD binaries merge at a rate at least 40 
times greater than the formation rate of stable mass-transfer AM~CVn systems, and so 
the majority must merge into single objects like R~CrB stars \citep{brown16b}.  The 
implication is that spin-orbit coupling in He+CO WD mergers is very weak 
\citep{marsh04}, or else nova-like outbursts during the initial phase of 
mass-transfer drive the systems into merger \citep{shen15}.

	Eclipsing double WD binaries are especially well-constrained systems.  Yet 
J0822 and the other 6 known eclipsing binaries do not contain a single ``normal'' 
0.6~\msun\ to 0.7~\msun\ WD.  While four eclipsing binaries have expected mass 
ratios of 1:1 to 1:2, three have extreme 1:5 mass ratios in tension with binary 
population synthesis models.  Finding and characterizing eclipsing WD binaries like 
J0822 is important for better understanding WD binary evolution and constraining 
the final outcome of WD binary mergers.

\acknowledgements

	We thank B. Kunk, E.\ Martin, and A.\ Milone for their assistance with 
observations obtained at the MMT Observatory.  This research makes use the SAO/NASA 
Astrophysics Data System Bibliographic Service.  This project makes use of data 
products from the Sloan Digital Sky Survey, which is managed by the Astrophysical 
Research Consortium for the Participating Institutions.  This work was supported in 
part by the Smithsonian Institution, and in part by the NSF and NASA under grants 
AST-1312678 and NNX14AF65G.

\facilities{MMT (Blue Channel Spectrograph), ARC (Agile Camera)}

\software{IRAF \citep{tody86, tody93}, RVSAO \citep{kurtz98}, JKTEBOP \citep{southworth04}}


\end{document}